\documentclass[article,twocolumn,superscriptaddress,floatfix]{revtex4-1}

\usepackage{graphicx}% Include figure files
\usepackage{amsmath}
\usepackage{bm}% bold math
\usepackage{color}
\usepackage{natbib}
\usepackage{appendix}

\begin{document}

\title{Optimal correlation order in super-resolution optical fluctuation microscopy}
\author{S.~Vlasenko$^1$, A.~B.~Mikhalychev$^1$, I.~L.~Karuseichyk$^1$, D.~A.~Lyakhov$^2$, D.~L.~Michels$^2$, D.~Mogilevtsev}
\affiliation{B. I. Stepanov Institute of Physics, National Academy of Sciences of Belarus, Nezavisimosti Ave. 68, Minsk 220072, Belarus;\\
$^2$Computer, Electrical and Mathematical Science and Engineering Division, 4700 King Abdullah University of Science and Technology, Thuwal 23955-6900, Kingdom of Saudi Arabia}

\date{August, 2020}

\begin{abstract}
Here, we show that, contrary to the common opinion, the super-resolution optical fluctuation microscopy might not lead to ideally infinite super-resolution enhancement with increasing of the order of measured cumulants. Using information analysis for estimating error bounds on the determination of point sources positions, we show that reachable precision per measurement might be saturated with increasing of the order of the measured cumulants in the super-resolution regime. In fact, there is an optimal correlation order beyond which there is practically no improvement for objects of three and more point sources. However, for objects of just two sources, one still has an intuitively expected resolution increase with the cumulant order.  
\end{abstract}

\maketitle

\section{Introduction}

Super-resolution optical fluctuation imaging (SOFI) is a simple, versatile  microscopy method popular for biological imaging. It provides the possibilities of  enhancement beyond the diffraction limit for both lateral and axial resolution, reduction of influence of background illumination, technical simplicity, and possibility to use rather low intensities of the field \cite{dertinger2009fast, sroda1a2020sofism,chen2017small,descloux2018combined,moser2019cryo,geissbuehler2012mapping}. It is suitable for super-resolution 3D imaging of living cells and subcellular structures \cite{dedecker2012widely,cho2013simple,dertinger2012sofi,geissbuehler2014live,girsault2016sofi,grussmayer2020spectral}. SOFI can be combined  with other microscopy methods, such as, for example, light-sheet fluorescence microscopy \cite{chen2016two}, image scanning microscopy \cite{sroda1a2020sofism},  and is easily integrated into  existing microscopic set-ups: for example, wide-field \cite{dertinger2009fast,grussmayer2020spectral}, confocal \cite{chen2015three}, total internal reflection fluorescence microscope \cite{dertinger2010superresolution, geissbuehler2012mapping,dedecker2012widely}.  SOFI functions by measuring high-order intensity correlations of randomly emitting fluorescent sources. Initially, SOFI was demonstrated with blinking quantum dots (QDs) \cite{dertinger2009fast}. Soon, it was applied using organic dyes \cite{dertinger2010superresolution}, fluorescent proteins \cite{dedecker2012widely,geissbuehler2014live,vandenberg2017effect}, and semiconducting polymer dots \cite{chen2017small,sun2019semiconducting, chen2017multicolor}. Later, SOFI was shown to work by speckled illumination with fluorophores without natural blinking  \cite{kim2015superresolution,yeh20163d}. 

SOFI seems to deliver a promise of potentially unlimited spatial resolution. By measuring temporal cumulants of the $n$th order, the original version of SOFI \cite{dertinger2009fast} offered resolution enhancement by the factor of ${\sqrt{n}}$. Later, the improvements of the initial algorithm  were suggested for reaching linear scaling of resolution with the cumulant order \cite{dertinger2010achieving}. Modifications of SOFI with structured illumination were even shown to increase possible resolution enhancement to $2n$ and even more \cite{Classen:s,zhao2017resolution}. 

However, despite rather intensive research on SOFI, practically obtained enhancement remains modest (generally, just few times increase of resolution beyond the diffraction limit \cite{PMID:30602772}). So, quite lot of research effort was spent in attempts to retrieve potentially infinite resolution enhancement predicted for SOFI by overcoming possible practical limitations in measuring high-order correlations, such as additional noise by imperfections of set-ups, finite size of detector pixels, emitters degradation, acquired image artefacts,  etc. \cite{dertinger2013advances,geissbuehler2012mapping,yi2019moments,vandenberg2016model,moeyaert2020sofievaluator,peeters2017correcting,vandenberg2017effect,zou2018high,geissbuehler2014live,stein2015fourier,yi2019moments,yi2020cusp,sun2019semiconducting}.  Some effort was also spent on enhancing temporal resolution, i.e., on reducing the number of raw images \cite{geissbuehler2014live,zeng2015fast,jiang2016enhanced,vandenberg2016model}.

However, up to now there were no works raising questions about the very possibility  of infinite resolution enhancement in SOFI, despite the fact that predicted resolution enhancement is actually based on quite empirical considerations. These considerations stem from the formal similarity of spatial cumulants and intensity distributions. So, the ``cumulants image'' looks like a conventional image but with much narrower point-spread functions (PSFs) (actually, the $n$th power of PSFs for the $n$th order cumulants). However,  this simple empirical understanding of resolution was already quite a time ago shown to be rather tricky and problematic (see, for example, the review on empirical resolution criteria \cite{denDekker:97}). 

Here, we implement information analysis for obtaining lower bounds on the errors of emitter's position estimation. We  show that a common intuitive approach to SOFI resolution works  only in the simplest case of  just two emitters. For three and more emitters with separations less than the diffraction limit, SOFI might bring no resolution enhancement beyond the certain (and not large) cumulants order. 

The outline of the paper is as follows. The second section describes  the idea of SOFI. The third section represents the model that we consider. The fourth section provides the information analysis procedure via the Fisher information and Cramer-Rao inequality. In the fifth section, we discuss how the informational content of the correlation functions measurements changes in the super-resolution and sub-resolution regimes depending on the correlation order. The sixth section analyses the informational content of cumulants and gives lower bounds for estimation errors. 

\begin{figure}[tp]
    \centering
    \includegraphics[width=\linewidth]{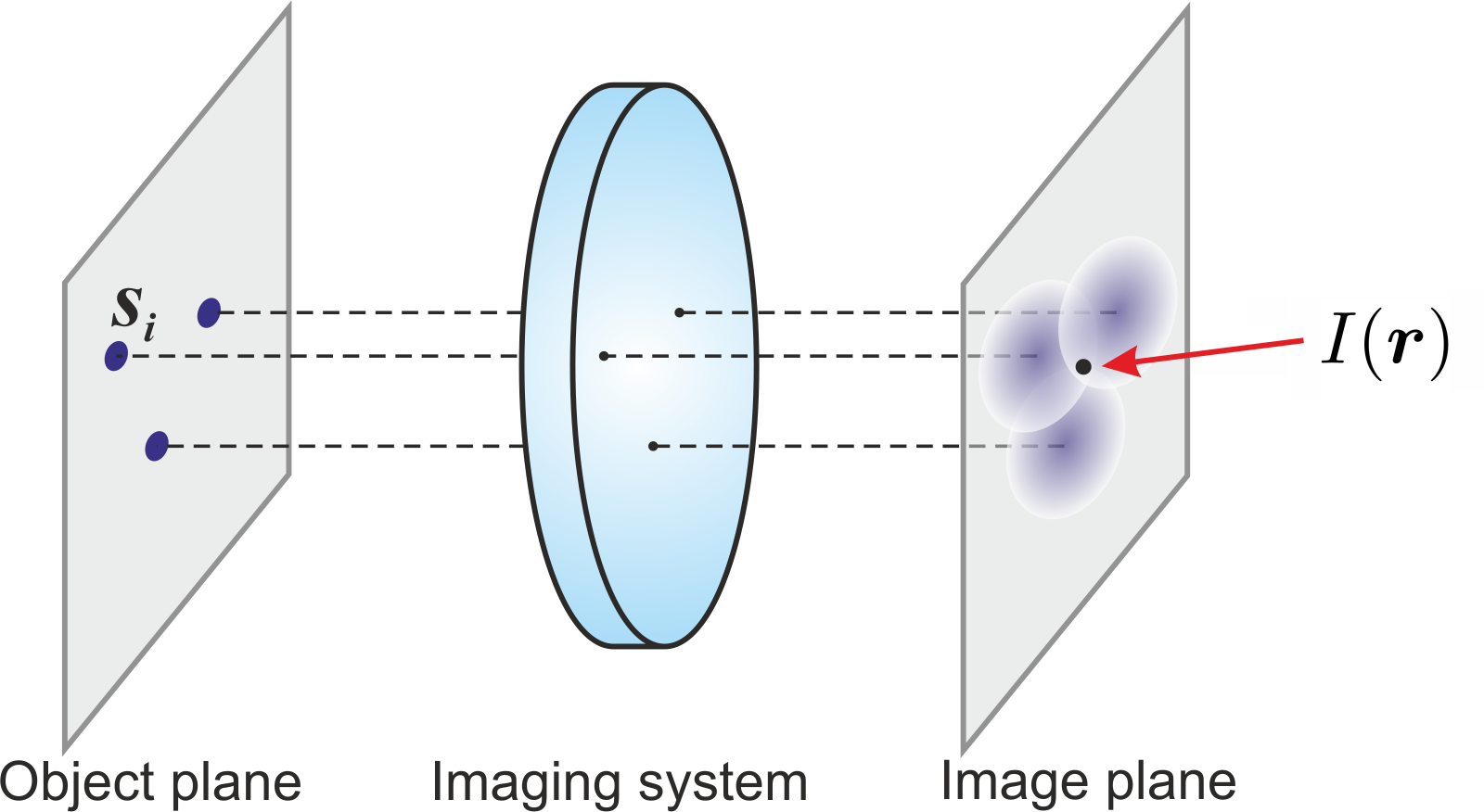}
    \caption{Basic scheme for imaging.}
    \label{fig:basic scheme SOFI}
\end{figure}

\begin{figure}[tp]
    \centering
    \includegraphics[width=\linewidth]{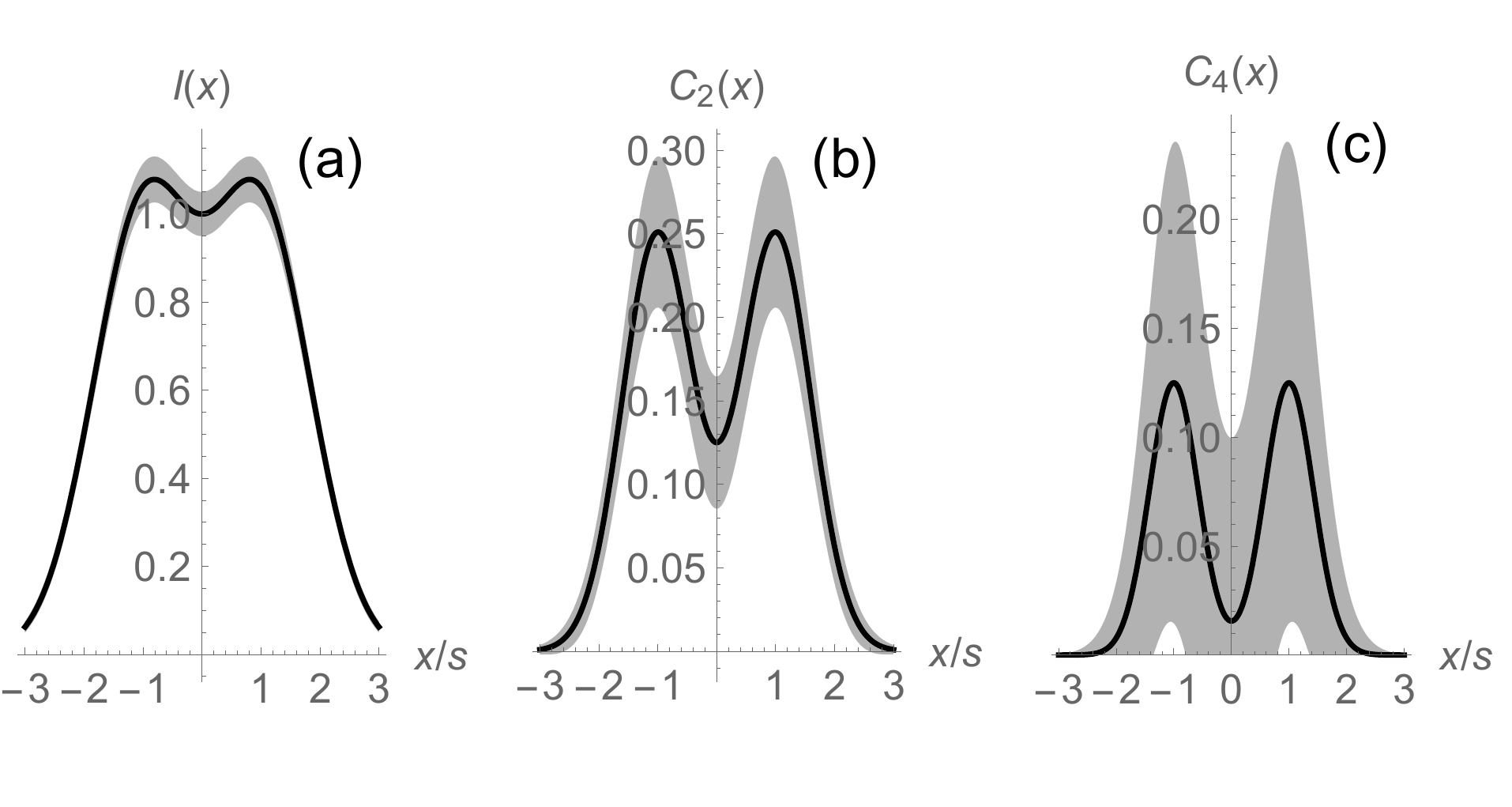}
    \caption{Images of two sources, placed at the points $s$ and $-s$, for intensity (a), 2nd (b) and 4th (c) order cumulants. The gray area indicates the fluctuations of the signal due to the shot noise. A Gaussian PSF with FWHM equal to the separation of the sources was used for modeling. The shown shot noise corresponds to 400 registered counts at the detectors placed at the intensity distribution maximums (see the Appendix A for more details).}
    \label{fig:noise}
\end{figure}

\section{Basic scheme}

Here, we briefly describe the basic concept of SOFI \cite{dertinger2009fast,dertinger2013advances}. Let us consider $M$ independent point stochastic sources, placed at the positions $\bm s_i$ of the object plane (Fig.~\ref{fig:basic scheme SOFI}) and imaged by an optical system with the PSF $h(\bm r - \bm s)$. The detected (average) intensity at the position $\bm r$ of the image plane can be represented as 
\begin{equation}
\label{intensity}
\langle I(\bm r) \rangle= \sum_{i=1}^{M} \left|h(\bm r - \bm s_{i})\right|^2 \langle I_{i}\rangle,
\end{equation}
where averaging is taken over the considered stochastic fluctuations of the sources (i.e. $\langle x \rangle$ denotes the expectation value of a stochastic process $x$); the random variable $I_{i}$ describes the contribution of the $i$th point source to the total intensity. If we can measure intensity moments up to some $n$th order, $\langle I^l(\bm r) \rangle$, $l=1\ldots n$, we can also construct cumulants $C^{(n)}(\bm r)$ from these moments. A cumulant of a sum of independent random variables is equal to the sum of individual cumulants. Therefore, 
\begin{equation}
\label{eqn:cumulant_ideal}
C^{(n)}(\bm r) = \sum_{i=1}^{M} \left|h(\bm r - \bm s_{i})\right|^{2n} c_{i}^{(n)},
\end{equation}
where $ c_{i}^{(n)}$ is the $n$th order cumulant of the random variable $I_{i}$. The expressions  (\ref{intensity}) and (\ref{eqn:cumulant_ideal}) are similar, and the ``cumulant image'' given by Eq.(\ref{eqn:cumulant_ideal}) looks like an ``intensity image'' (\ref{intensity}) with PSFs effectively raised to the power $n$. Thus, resolution increase of $\sqrt{n}$ for $n$-order ``cumulant image'' was surmised \cite{dertinger2009fast}. In Fig.~\ref{fig:noise}, one can see an illustration for this resolution enhancement for just two point sources with Gaussian PSFs for the ``intensity image'' (a), the second (b) and the fourth (c) order "cumulant image". 

Of course, in practice it is not straightforward to get such visible enhancement. First of all, real measurements are of finite duration, and one needs to approximate mathematical expectations by frequencies calculated for finite sets of data. One encounters shot noise, which is amplified when calculating cumulants (Appendix A). Additionally, the simple expression (\ref{eqn:cumulant_ideal}) does not hold any more for estimates of cumulants, based on finite data sets (i.e. cross-terms produced by different sources vanish only in average, but remain non-zero for a particular realization; see Appendix A). For the same measurement duration (number of runs), higher-order cumulants are more noisy than lower-order ones. This situation is  illustrated in Fig.~\ref{fig:noise} for the  ``intensity image'' (a), the second (b) and the fourth (c) order ``cumulant image'' built from the data collected during the same time interval  (error bars in dependence on the position are given by the gray shaded areas).  

Besides, for weak sources and detectors with finite efficiency, the absolute values of cumulants rapidly decrease with the growth of their order $n$. Thus, the resolution enhancement can be also spoiled for a number of reasons mentioned in the Introduction, such as detectors imperfections and additional noise, emitters bleaching or other degradation of sources, image artefacts,  etc. \cite{dertinger2013advances,geissbuehler2012mapping,yi2019moments,vandenberg2016model,moeyaert2020sofievaluator,peeters2017correcting,vandenberg2017effect,zou2018high,geissbuehler2014live,stein2015fourier,yi2019moments,yi2020cusp,sun2019semiconducting}. 

The main message of our paper is that even in the absence of imperfections, for ideally obtained ``cumulant image'' satisfying Eq. (\ref{eqn:cumulant_ideal}), there are still resolution limits dictated by the very nature of the SOFI. The empirical ``resolution enhancement'' picture, illustrated by solid lines in Fig.~\ref{fig:noise}, does not hold for more complicated objects than just two point sources as soon as one takes into account inevitable shot noise. Increasing the cumulants order might not lead to the actual decrease of error in estimation of sources' locations, if these sources are close enough. 

To get deeper understanding of the dependence of achievable resolution on the order of cumulants, one needs to go beyond  empirical considerations of the ``PSF narrowing'' and to consider the  informational content of the obtained images. The purpose of the current contribution is to demonstrate that analysis of the Fisher information provides a finite value of the optimal correlation order $n$ for objects of three and more sources in the super-resolution regime, in contrast to the infinite resolution increase predicted from empirical considerations.

\section{Model}

We demonstrate the analysis of information content of the measured data with the help of the following simple model, where the imaged object is composed of weak non-Gaussian sources. Let the object consist of $M$ independent point sources, situated at the positions $\bm s_i$ at the object plane. The light, emitted by the sources, is mapped onto the image plane by an optical system with the transfer function (PSF) $h(\bm r - \bm s)$. Further, for simplicity's sake, we assume the function to have a Gaussian shape:
\begin{equation}
\label{eqn:PSF}
    h(\bm r -\bm s)=\frac{1}{\sqrt{\pi}w}\exp\left[-\frac{ (\bm r-\bm s)^2}{w^2}\right],
\end{equation}
where $w$ characterizes the width of the PSF (its FWHM equals $2 w \sqrt{\log 2}$).

We assume a simplistic model of the point sources as single-mode ones. For such sources, the positive-frequency field operator in the object plane can be written as
\begin{equation}
E_0^{(+)}(\bm s)\sim \sum_{i=1}^M \delta^{(2)}(\bm s- \bm s_i)a_i,
\end{equation}
where $a_i$ is the annihilation operator for the field produced by the $i$th source.

To fix a particular state of emitted light for further modeling, we assume that the density matrix of the field mode, corresponding to the $j$th source, is 
\begin{equation}
\label{eqn:state}
    \rho_j=\xi_j\overline{|\alpha_{j}\rangle \langle \alpha_{j}|}+ (1 - \xi_j) |0\rangle \langle 0 |,
\end{equation}
where $\xi_j$ is the probability of the source being in its ``bright'' state and 
\begin{equation}
    \overline{|\alpha_j \rangle \langle \alpha_j|} = \frac{1}{2 \pi} \int_0^{2 \pi} d \varphi |\alpha_j e^{i \varphi} \rangle \langle \alpha_j e^{i \varphi}|
\end{equation}
is a phase-averaged coherent state with the amplitude $\alpha_j$. We take that the source is realized in such a way that its state switches randomly from the ``bright'' state to the vacuum during average switching time $\tau_0$, and the mixture state  (\ref{eqn:state}) is observed for sufficiently long time $T\gg \tau_0$. Notice that to simplify the discussion, for the major part of our consideration we assume our sources to be identical, $\alpha_j=\alpha$, $\xi_j=\xi$.

Regardless of trivial behavior of cumulants for the ``bright'' state with Poissonian photon statistics, the overall quantum state, averaged over both ``bright'' and ``dark'' regimes of the emitter, is a non-Gaussian state with non-trivial dependence of cumulant values on their order.

We assume that our detectors measure correlation functions of the field at the image plane. The $n$th order single-time correlation function at the image plane reads
\begin{multline}
\label{eqn:Gn}
G^{(n)}(\bm r_1, \ldots, \bm r_n)= \\
\left\langle E^{(-)}(\bm r_1) \ldots E^{(-)}(\bm r_n) E^{(+)}(\bm r_n) \ldots E^{(+)}(\bm r_1) \right\rangle,
\end{multline}
where the positive-frequency field operator in the image plane is expressed in terms of the field $E_0^{(+)}(\bm s)$ in the object plane in the following way:
\begin{equation}
E^{(+)}(\bm r)=\int d^2 \bm s\,E_0^{(+)}(\bm s)h(\bm r -\bm s),
\end{equation}
and
\begin{equation}
E^{(-)}(\bm r)=\left[E^{(+)}(\bm r)\right]^+
\end{equation}
is the negative-frequency field operator.

As it is mentioned above, for the ``basic'' SOFI scheme considered in the previous Section, the $n$th order single-point single-time cumulant $C^{(n)}(\bm r)$ can be built from intensity moments up to the $n$th order: $\langle I^l(\bm r) \rangle$, $l=1\ldots n$. For example, the second-order cumulant is expressed as $C^{(2)}(\bm r) = \langle I^2(\bm r) \rangle - \langle I(\bm r) \rangle^2$.  The $l$th order intensity moment, $\langle I^l(\bm r) \rangle = \langle \left[E^{(-)}(\bm r) E^{(+)}(\bm r) \right]^l \rangle$, is defined by the measured single-point correlation functions $G^{(n)}(\bm r) \equiv G^{(n)}(\bm r, \ldots, \bm r) = \langle {:} \left[E^{(-)}(\bm r) E^{(+)}(\bm r) \right]^l {:} \rangle$, up the $l$th order. Here, ${:}X{:}$ denotes normal ordering of the field creation and annihilation operators inside the operator $X$. Thus, the $n$th order cumulant comprises the information contained in the correlation functions up to the $n$th order: $G^{(l)}(\bm r)$, $l=1 \ldots n$.

 Further, we will pay more attention to the correlation functions themselves, bearing in mind that the $n$th order cumulant cannot contain more information than the amount provided by the correlation functions $G^{(l)}(\bm r)$, $l=1 \ldots n$, in total.

Eqs.~(\ref{eqn:state}),~(\ref{eqn:Gn}) yield the following result for the $n$th order correlation function:
\begin{eqnarray}
    \label{eqn:Gn_final}
    G^{(n)}(\bm r) = \sum_{\begin{array}{c}
         \scriptstyle n_1,\ldots,n_M:  \\
         \nonumber
         \scriptstyle n_1 + \cdots + n_m = n 
    \end{array} }
    \frac{n!}{n_1!\cdots n_M!} \\{} \times \prod_j |\alpha_j|^{2n_j}  \left|h(\bm r - \bm s_j)\right| ^{2 n_j} \left( \xi_j  + \delta_{n_j0} (1 - \xi_j) \right).
\end{eqnarray}
For high orders $n$ of the correlations, $n > M$, it is also convenient to rewrite this expression as
\begin{multline}
    \label{eqn:Gn_final2}
    G^{(n)}(\bm r) =\Bigl( \prod_i \xi_i\Bigr) \Bigl(\sum_i q_i \Bigr)^n \\{} + \sum_j (1 - \xi_j) \Bigl(\prod_{i: i\ne j} \xi_i\Bigr) \Bigl( \sum_{i: i\ne j} q_i \Bigr)^n \\
    {} + \sum_{j,k:j>k} (1 - \xi_j)(1 - \xi_k) \Bigl(\prod_{i: i\ne j, i\ne k} \xi_i\Bigr) \Bigl( \sum_{i: i\ne j, i\ne k} q_i \Bigr)^n \\{} + \cdots,
\end{multline}
where $q_i = |\alpha_i|^2 |h(\bm r - \bm s_i)|^2$.

We emphasize that our simple model of the sources is perfectly suitable for application of the empirical SOFI considerations described in the Section II. Recording high-order spatial correlation functions at the image plane and building ``cumulant image'' from them gives one PSF narrowing essential for the SOFI.

\begin{figure*}[tp]
    \centering
    \includegraphics[width=\linewidth]{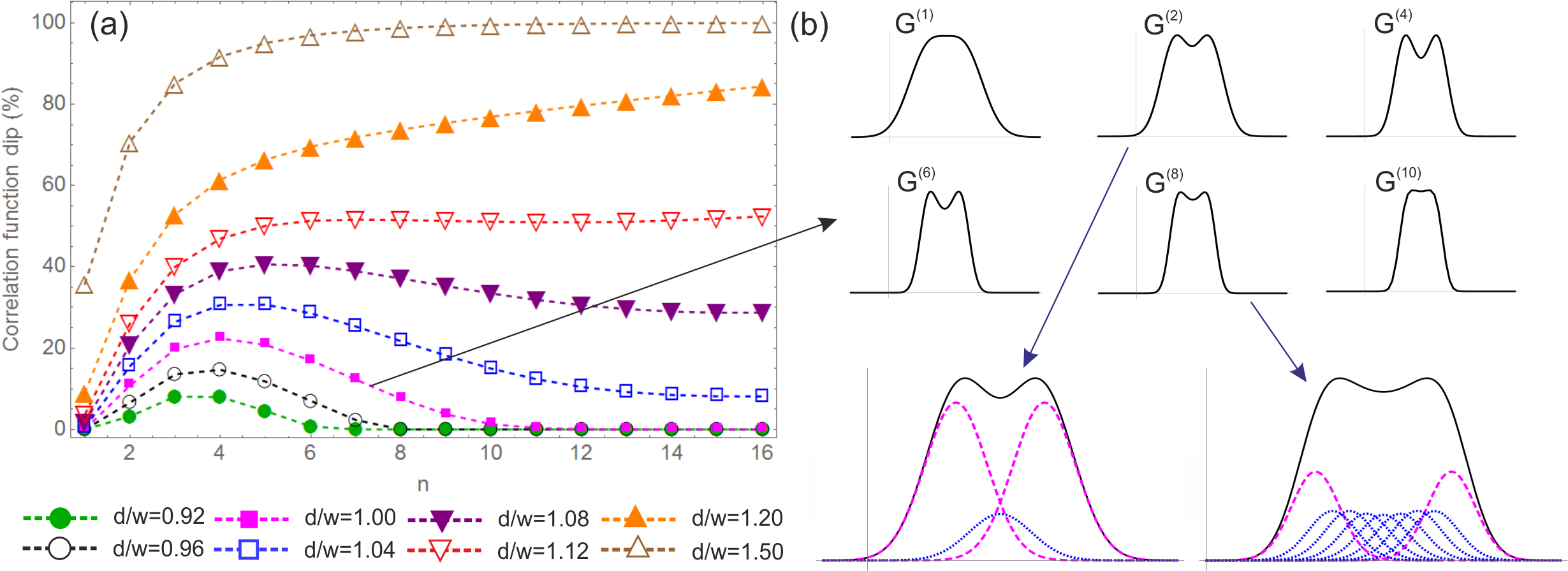}
    \caption{The dependence of the dip between the maxima of correlation functions on the image plane produced by a two-source object on the correlation order (a).  Different curves correspond to various  distances  between the sources. The panel (b) depicts  correlations functions of different orders for the fixed distance between the sources ($d/w=1$). The correlation functions are given by Eqs.(\ref{eqn:Gn_final}), (\ref{eqn:Gn_final2}). The amplitude $\alpha = 0.3$ of the coherent state generated in the ``bright'' regime of the source and the probability of that regime $\xi = 0.4$ are the same for the two sources. The lower row of the panel (b) shows a more detailed view of the 2nd and 8th order correlation functions, decomposed according to Eq. (\ref{eqn:Gn_2sources}): dashed lines represent the main maxima, becoming narrower with the growth of the correlation order; dotted lines show the contribution of the additional terms, reducing the contrast and decreasing the achievable resolution.}
    \label{fig:corr function gap}
\end{figure*}

\section{Data acquisition and Fisher information} 

We assume a possibility of an arbitrarily long data acquisition, and limit ourselves with analyzing only single-point  and single-time  correlation functions. Also, for simplicity, we assume that the shot noise of the quantities $G^{(n)}(\bm r)$ and $G^{(n)}(\bm r')$ is independent for $\bm r \ne \bm r'$. This condition is fulfilled, for example, for the measurements in the scanning regime. 

The cumulant of $n$th order is composed of correlation functions of orders from $1$ to $n$. We assume that all these correlation functions are measured independently at the discrete set of image plane points $\{\bm r_j\}$. For such set of the $n$th order correlation functions, one can introduce the normalized set of quantities
\begin{equation}
    \label{eqn:Gn_normalized_discrete}
    p_j^{(n)} = \bar G^{(n)}(\bm r_j) = \frac{G^{(n)}(\bm r_j)}{\sum_k G^{(n)}(\bm r_k)},
\end{equation}
where the sum is taken over the set of the considered detection points.
By the definition, the quantities $p_j^{(n)}$ sum up to unity and can be interpreted as probabilities of the possible detection outcomes: $p_j^{(n)}$ describes the conditional probability of detecting $n$ photons at the particular point $\bm r_j$, if it is known that $n$ photons have been detected at any of the detection points $\{\bm r_i\}$. Notice that taking the normalized form (\ref{eqn:Gn_normalized_discrete}), one eliminates the effect of the finite detection efficiency, because it enters both the numerator and the denominator of Eq. (\ref{eqn:Gn_normalized_discrete}).  

For the introduced set of probabilities for the measured correlation functions of $n$th order, the Fisher information matrix can be calculated in the following standard way \cite{doi:10.1142/S0219749909004839}: 
\begin{equation}
F_{\mu\nu}^{(n)}=\sum_j \frac{1}{p_j^{(n)}}\frac{\partial p_j^{(n)}}{\partial\theta_\mu}\frac{\partial p_j^{(n)}}{\partial\theta_\nu},
\label{eqn:Fisher_matrix_def}
\end{equation}
where $\{\theta_\mu\}$ is the set of parameters of interest, describing the investigated configuration of the sources. Further, we assume that we are interested in the positions of the sources, the set of the parameters being $\{x_1, \ldots, x_M\}$ for a 1-dimensional configuration or $\{x_1, y_1, \ldots, x_M, y_M\}$ for a 2-dimensional object.

According to the Cram{\'e}r-Rao inequality, the variance for an unbiased estimator for the parameter $\theta_\mu$ is bounded by the corresponding diagonal element of the inverse of the Fisher information matrix \cite{doi:10.1142/S0219749909004839}:
\begin{equation}
    \label{eqn:Cramer-Rao}
    \operatorname{Var}(\theta_\mu) \ge \frac{1}{N} \left([F^{(n)}]^{-1}\right)_{\mu \mu},
\end{equation}
where $N$ is the number of detected events. The overall quality of the unknown parameters reconstruction can be quantified by the sum of the parameters' variances:
\begin{equation}
    \label{eqn:Delta2}
    \Delta^2 = \sum_\mu \operatorname{Var}(\theta_\mu) \ge \frac{1}{N} \operatorname{Tr}\left( [F^{(n)}]^{-1}\right).
\end{equation}

Assuming that the number of registered events is the same for the measurements of all the correlation functions required to build a $n$th order cumulant and taking into account the measurements' independence, one can use the additivity of the Fisher information. Therefore, the informational content of the cumulant $C_n(\bm r)$ can be estimated from the sum of the Fisher matrices, describing the corresponding correlation functions:
\begin{equation}
    \label{eqn:Fisher_matrix_total}
    F_{\mu \nu}^{(n,\Sigma)} = \sum_{m=1}^n F_{\mu \nu}^{(m)}.
\end{equation}
The lower bound of the total error of the parameters reconstruction on the base of the cumulant $C_n(\bm r)$ is, therefore, given by the following expression:
\begin{equation}
    \label{eqn:Delta2_total}
    \Delta_n^2 = \sum_\mu \operatorname{Var}(\theta_\mu) \ge \frac{1}{N} \operatorname{Tr}\left[ \left(F^{(n,\Sigma)}\right)^{-1}\right].
\end{equation}
Further, we will compare the informational content of different correlation orders per one detected event, i.e. take $N = 1$.

\section{Correlation functions}

The data acquisition scheme described in the previous Section is really favorable to the experimenter inclined to demonstrate infinite resolution enhancement with increasing cumulants order (no additional noise, long acquisition time, independent measurements of correlations at each point, etc.). However, even in such conditions informational content of the measurement might be actually dropping with the increase of the correlation order. 

Notice that the behavior of correlation functions even for a simple two-source object hints to rather nontrivial properties of the image inferred from the higher-order correlations. Fig.~\ref{fig:corr function gap} shows how the single-point $n$th order correlation function might behave with increasing of the order $n$. For a relatively large value of the distance between the sources, $d$, normalized by the width of the PSF, $w$, a dip between the maxima corresponding to the positions of the sources, increases with the growth of the correlation order as one would expect. However, in the super-resolution regime (for $d$ close to $w$ and less), this dip behaves in a quite non-monotonous way. There is a finite correlation order corresponding to the maximum contrast in Fig.~\ref{fig:corr function gap} (b). 

To understand the reasons for such behavior, let us consider Eq. (\ref{eqn:Gn_final2}) for two identical sources:
\begin{equation}
    \label{eqn:Gn_2sources}
    G^{(n)}(\bm r) = \xi (q_1^n + q_2^n) + \xi^2 \sum_{m = 1}^{n-1} \binom{n}{m} q_1^m q_2^{m-n}.
\end{equation}
The first term at the right-hand side of Eq. (\ref{eqn:Gn_2sources}) describes two maxima, formed by peaks at the positions $\bm r = \bm s_1$ and $\bm r = \bm s_2$ (Fig. \ref{fig:corr function gap}(b), lower row). As expected, the peaks become narrower with the growth of the correlation order $n$, because $q_i^n \propto |h(\bm r - \bm s_i)|^{2n}$. The term $q_1^m q_2^{m-n} \propto |h(\bm r - \bm s_1)|^{2m} |h(\bm r - \bm s_2)|^{2(n-m)}$ produces a peak with a maximum at the position $\bm r = (m / n) \bm s_1 + ((n-m) / n) \bm s_2$ in between the two main maxima. Therefore, the terms with $m = 1$, \ldots, $n-1$ reduce the dip between the maxima and decrease the image contrast. The total weight of such terms, $2^n - 2$, increases exponentially with the growth of the correlation order $n$. The interplay of the two discussed effects results in the presence of the optimal (the most informative) correlation order, yielding the image with the highest contrast.  
The peaks, causing the contrast decrease, are formed due to non-zero overlap of $h(\bm r - \bm s_1)$ and $h(\bm r - \bm s_2)$ and effectively vanish for classically resolved sources ($d/w >1$).

\begin{figure*}[tp]
    \centering
    \includegraphics[width=0.7\linewidth]{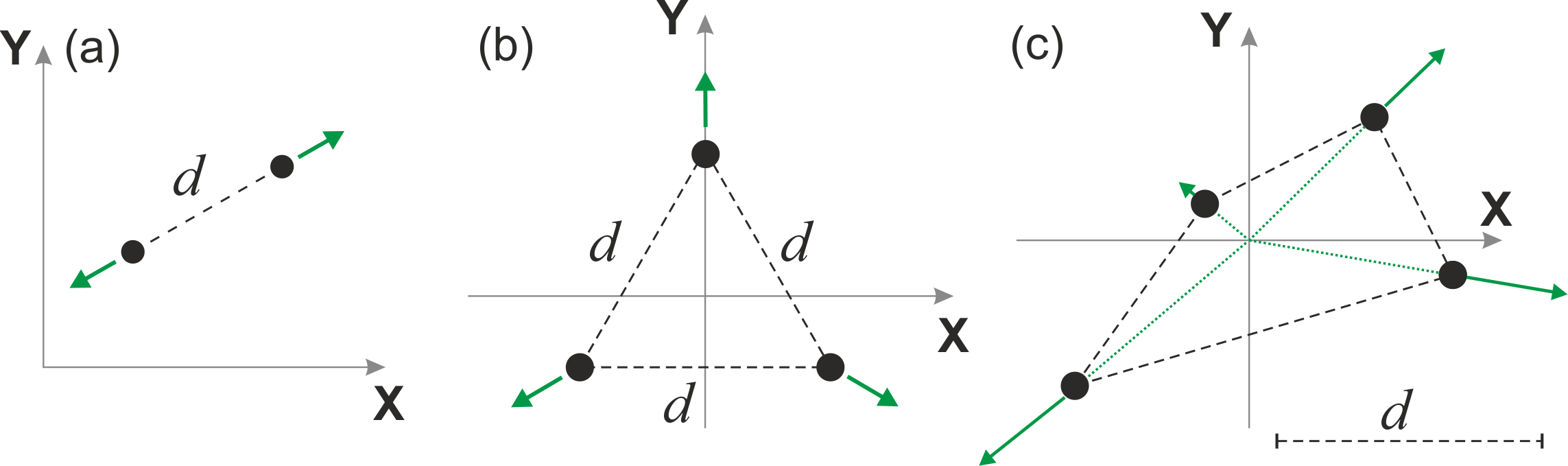}
    \caption{Schematic illustration of the object configurations with 2 (a), 3 (b) and 4 (c) sources for the two-dimensional case. Arrows show the direction of the sources displacement as the distance between the sources is scaled: (a) along the line, connecting the sources, (b) from the center of the triangle, (c) from the origin of the coordinate system, so that an asymmetric configuration of the sources is maintained, with the coordinates of the sources varying as $\{s_{1x}, s_{1y}\}=\{-0.2d, 0.15d\}$, $\{s_{2x}, s_{2y}\}=\{0.465d, 0.45d\}$, $\{s_{3x}, s_{3y}\}=\{0.85d, -0.25d\}$, $\{s_{4x}, s_{4y}\}=\{-0.73d, -0.55d\}$. For the configurations with 2 (a) and 3 (b) sources, $d$ is the distance between the sources.}
    \label{fig:configuration 2D}
\end{figure*}

Somewhat similar behavior is also observed in the informational content of the measured correlation functions.  We demonstrate it for several simple configurations with 2, 3 and 4 sources both in 1D and 2D cases (the latter case is depicted in Fig.~\ref{fig:configuration 2D}). The 1D case is represented by just 2, 3 or 4 equidistant point sources on the line. 

To make the results more illustrative, we choose just a single ``collective'' parameter, $d$, to characterize the scaling of the sources configuration. Nevertheless, we consider a multi-parametric problem of object inference. No information about the relations between the coordinates of the sources and the ``collective'' scale $d$ (similar to the ones shown in Fig. \ref{fig:configuration 2D}) is assumed to be available for the observer. The dimension of the Fisher matrix equals to the number of parameters: $M$ for the 1-dimensional case and $2M$ for the 2-dimensional one, where $M$ is the number of sources, representing the object ($M = 2$, $3$, and $4$). To analyze the optical resolution, we consider a set of problems for each configuration of the sources by effectively re-scaling it: we vary the scale parameter $d$ (typical distance between adjacent sources) while keeping the shape of the configuration constant and characterize the scale by the dimensionless parameter $d / w$, where $w$ (see Eq.~(\ref{eqn:PSF})) is the PSF width.

\begin{figure}[tp]
    \centering
    \includegraphics[width=0.45\textwidth]{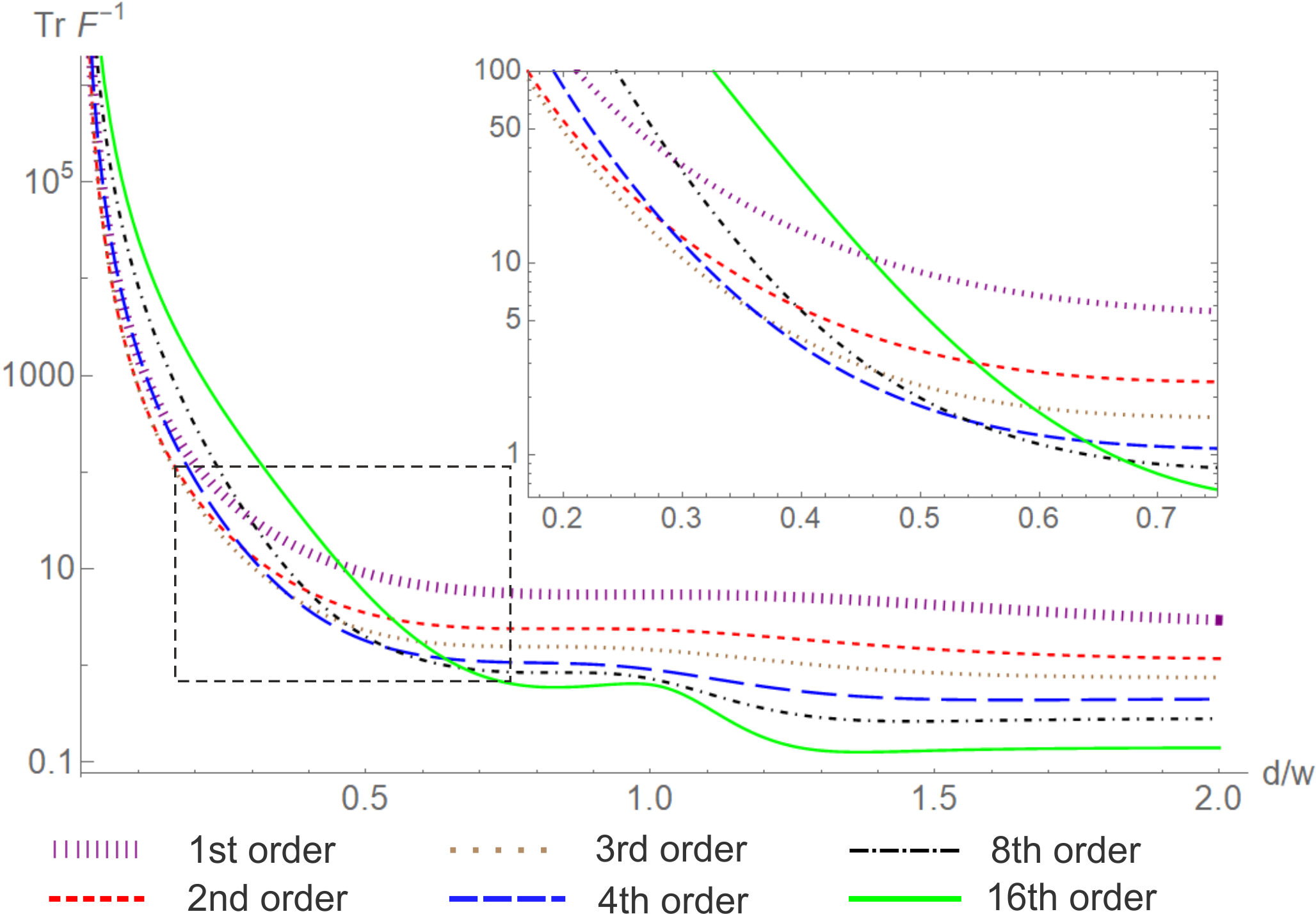}
    \caption{The dependence of the total reconstruction error, estimated as the trace of the inverse of Fisher information matrix, on the normalized distance between adjacent sources for 3 sources in the 1-dimensional case. The curves correspond to different orders of the analyzed correlation functions. The inset shows an enlarged part of the plot with intersections of the curves. The amplitude $\alpha = 0.3$ of the coherent state generated in the ``bright'' regime of the source and the probability of that regime $\xi = 0.4$ are the same for all the sources. The probabilities (\ref{eqn:Gn_normalized_discrete}) were evaluated at the discrete set of image plane points, represented by a grid with the step $\Delta x = 0.02$ covering the whole region, where the signal has essentially non-zero values (practically, it is the interval $ x \in (x_i-2w, x_j+2w)$, with $x_i$ and $x_j$ being the coordinates of the most left and the most right sources of the configuration).}
    \label{fig:error vs scale}
\end{figure}

Fig.~\ref{fig:error vs scale} shows typical behavior of the lower bound on the total object reconstruction error per measurement (i.e. the trace of the inverse Fisher matrix) with the example of the object composed of three identical sources on a line. Measurements of correlation functions of different orders are considered. For Fig.~\ref{fig:error vs scale}, each of the point sources generates the state (\ref{eqn:state}) with $\alpha = 0.3$ and $\xi = 0.4$.

The results are similar to the intuitive picture described in  Fig.~\ref{fig:corr function gap}. For $d/w$ close to unity and larger (regime of ``classical'' resolution), the trace of the inverse Fisher matrix diminishes with the growth of the correlation order, yielding  increase of the resolution. For smaller $d / w$ (super-resolution regime), the situation is opposite. The lower bound for the reconstruction error becomes larger for the higher-order correlations (see the inset in Fig.~\ref{fig:error vs scale} ), and, for example, for $d / w$ of about $0.5$ it is more useful to take first- or second-order correlations to infer the object parameters. 

The error bounds behavior in dependence on $d / w$ and correlation order, depicted in  Fig.~\ref{fig:error vs scale}, is not just a feature of a particular object. In Appendix B, we have calculated error bounds for several 1D and 2D objects (such as the 2D ones shown in Fig.~\ref{fig:configuration 2D}), and found the same qualitative features.  The informational content of the correlation function can be indeed dropping with an increasing correlation order in the super-resolution regime. 

\begin{figure*}[tp]
    \centering
    \includegraphics[width=0.9\textwidth]{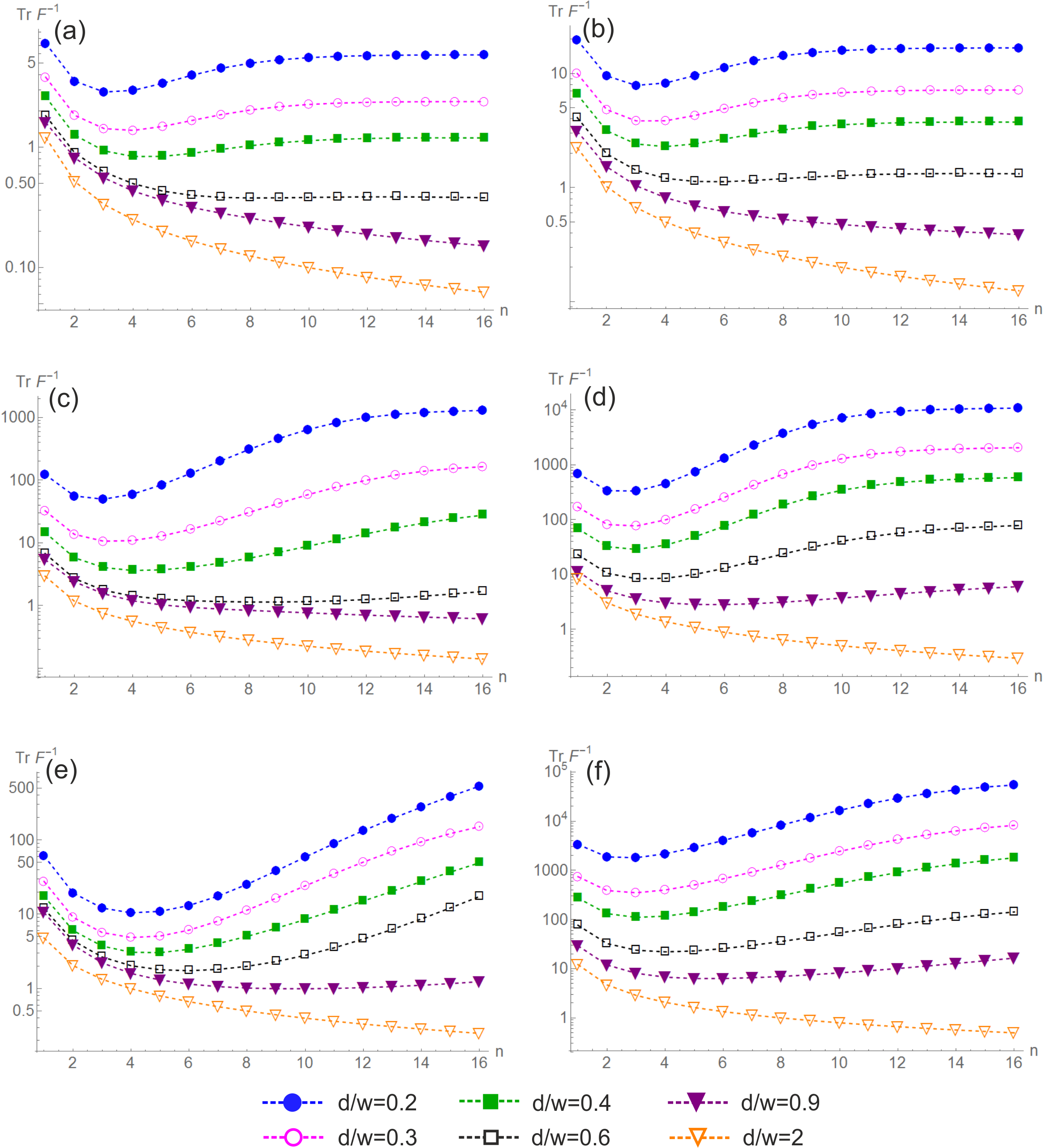}
    \caption{The dependence of the total reconstruction error, estimated as the trace of the inverse of Fisher information matrix, on the order of the analyzed correlation function for 2 (a), (b), 3 (c), (d) and 4 (e), (f) sources in the 1-dimensional (left) and 2-dimensional (right) cases. The curves correspond to different regimes of imaging, determined by the normalized distance between the sources $d / w$. The parameters $\alpha = 0.3$ and $\xi = 0.4$ are the same for all the sources. The set of detection positions was taken as for Fig.\ref{fig:error vs scale}; in 2D case, a square grid, satisfying the same conditions, was used for both axes.}
    \label{fig:error vs order}
\end{figure*}

This feature points to the conclusion about the existence of an optimal correlation order for the object inference.  Fig.~\ref{fig:error vs order} demonstrates it for different 1D and 2D objects. For all the considered cases, one can see that in the super-resolution regime (for $d/w$ being $0.5$ and lower), the most informative correlation function has a comparatively low order (4th at most).

In the Appendix B, we have also calculated the trace of the inverse Fisher matrix for the objects with identical sources, but for different probabilities of the ``bright'' state than the one considered in  Fig.~\ref{fig:error vs order}. Also, the calculations were performed for the objects composed of the sources with different amplitudes of the ``bright'' state. 
 
The general tendency remains the same: in the super-resolution regime correlation functions of comparatively low order are the most informative for the object reconstruction. Moreover, for ``brighter'' sources spending more time in the emitting state, the lowest-order correlation function (i.e. the intensity) might be the best for inferring object parameters.  

Curiously, the similar tendency was noticed even for thermal sources: informational analysis akin to the one described above, it was shown that for determination of the spatial characteristics of the extended thermal source measurement of the lower-order intensity correlation functions can be better than the measurement of the higher-order ones \cite{kok2015}. 

\section{SOFI errors}

The fact of having informational content of the correlation functions dropping with the increase of their order leads to possible limitation of the SOFI. It means that, in the super-resolution regime  after some correlation order, the information increase with the growth of the cumulant order might be negligibly small. In a somewhat paradoxical way, empirically surmised  ``squeezing'' of the PSF might not correspond to the actual resolution enhancement as the possibility to infer the object parameters more precisely. 

\begin{figure*}[tp]
    \centering
    \includegraphics[width=0.9\textwidth]{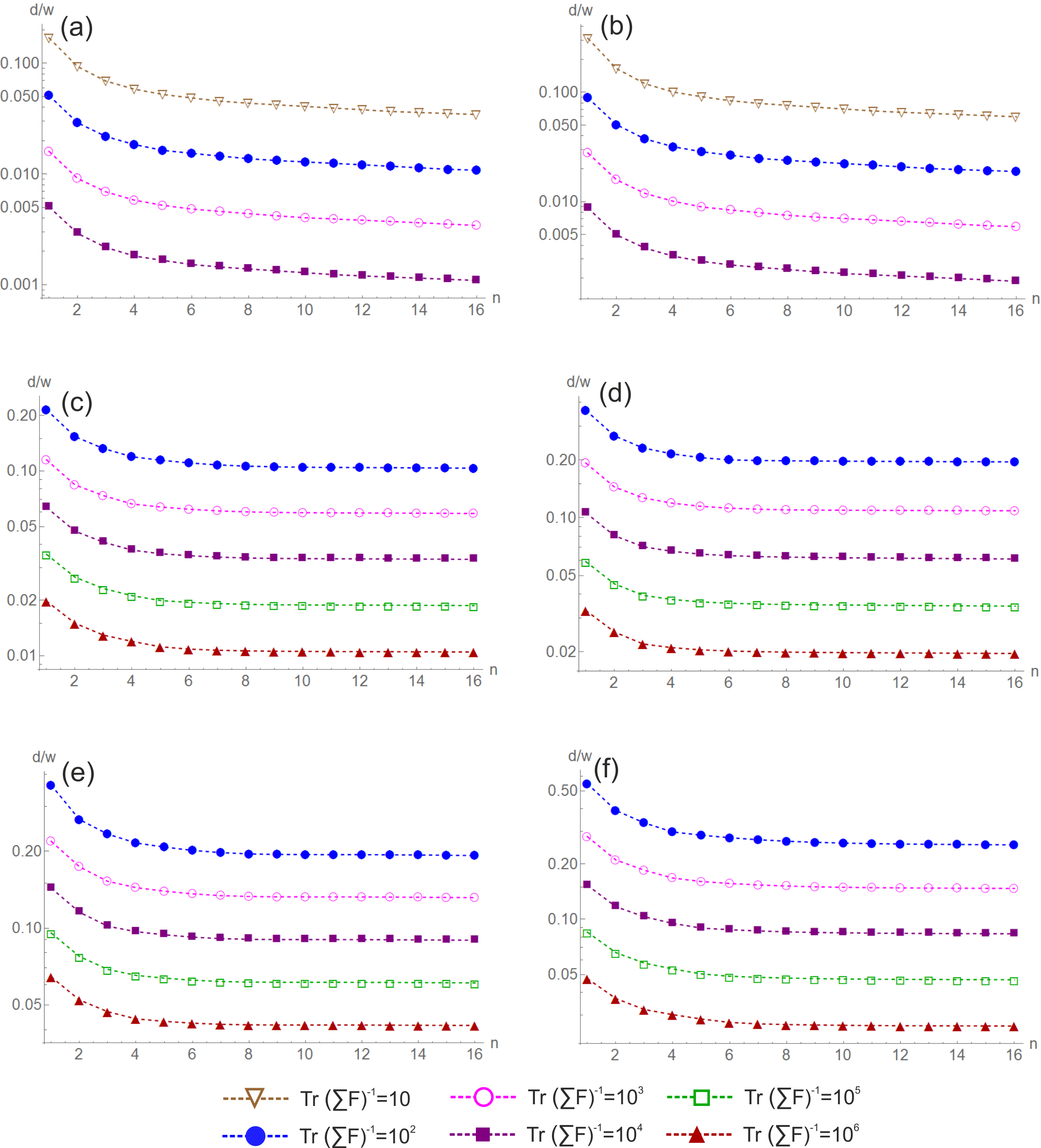}
    \caption{Estimate of the achievable resolution for the inference of the object parameters by considering cumulants of different orders. The resolution is quantified by the minimal normalized distance $d/w$ between the sources, for which the total reconstruction error $\operatorname{Tr}F^{-1}$ is smaller than the threshold value (indicated in the plot legend). The modeling is performed for 2 (a), (b), 3 (c), (d) and 4 (e), (f) sources in the 1-dimensional (left) and 2-dimensional (right) cases. The curves correspond to different acceptable values of the total reconstruction error per a measurement run (i.e. to the experiments with different numbers of measurement runs, and therefore, different acquisition times). For $n$th order cumulant, the Fisher information is summed over all the orders of correlation functions up to and including $n$  --- Eq.~(\ref{eqn:Fisher_matrix_total}). The parameters $\alpha = 0.3$ and $\xi = 0.4$ are the same for all the sources. The sources detection positions were taken as for Fig.\ref{fig:error vs order}. }
    \label{fig:total fisher info}
\end{figure*}

Fig.~\ref{fig:total fisher info} illustrates this situation showing the dependence of the resolution estimate on the order of the analyzed cumulant (it is worth noticing that the prediction is rather optimistic in relation with the real SOFI,  according to Eq.~(\ref{eqn:Delta2_total})). 

Fig.~\ref{fig:total fisher info} uncovers a possible reason for the established opinion of potentially ``infinite'' resolution achievable with the SOFI. As the panels \ref{fig:total fisher info} (a,b) show, the case of the object composed of just two point sources stands apart from the cases of more complicated objects. For the object of two sources, one has an expected decrease of the lower bound of the error with the growth of the correlation order. For more complicated objects, the situation is different. Going beyond $n \sim 6$ does not provide any significant advantages for the considered set of 1D and 2D objects. 

Here, one can draw a parallel with the recent lively discussion on the "dispelling" the infamous "Rayleigh curse" with the imaging of two incoherent point sources: when just intensity image is registered, the Fisher information tends to zero with the distance between the sources tending to zero not allowing the sources to be resolved.  One can devise a measurement to make the Fisher information non-zero even for the zero distance between the sources \cite{PhysRevX.6.031033,PhysRevLett.117.190802}. However, it is not possible for objects composed of three and more sources. In this case, the lower error bound always tends to infinity with reducing the object scale \cite{PhysRevA.99.012305,PhysRevA.99.013808}.

\section{Conclusions}

In this work, we demonstrated that even for idealized arbitrarily long, perfect independent measurements, SOFI  might not bring infinite lowering of statistical errors per measurement for objects more complicated than just two point sources. With the examples of simple objects, composed of three and four point sources, we showed that the lower bound for the total error of object parameter estimation (namely, positions of the sources) might tend to the constant value with increasing of the cumulants order. So, measuring intensity correlation functions beyond the fifth or sixth order brings no improvement of the total error bound by the SOFI. Notice that such phenomenon takes place exactly in the parameter region where one seeks to get a resolution gain, i.e., beyond the conventional diffraction limit. For larger object sizes (classically resolved ones), increasing correlation order brings resolution improvements, as it is intuitively expected. 

So, our data confirm the already established opinion of the SOFI being able to bring only moderate (less than three times in our examples with three and four sources) improvements over the diffraction limit in the realistic microscopic scenarios. Just few times over the diffraction limit seems to be a maximal gain that one should expect obtaining via the SOFI. 

However, one should emphasize that our conclusion holds only for the standard SOFI scenario with independent non-Gaussian point sources, and not for other imaging methods based on the analysis of the high-order field correlations, for example, the SOFI version with the structured illumination \cite{Classen:s,zhao2017resolution}. 

All the authors acknowledge financial support from the King Abdullah University of Science and Technology (grant 4264.01), S. V., A. M., and D.M. also acknowledge support from the EU Flagship on Quantum Technologies, project PhoG (820365).

\appendix

\section{Statistical errors for different cumulants}

If $I_i(t)$ is fluctuating intensity of the $i$th emitter and $h(\bm r - \bm s)$ is the point-spread function (PSF) of the imaging optics, the $n$th order single-point single-time cumulant of the registered signal will be expressed in the following way \cite{dertinger2009fast,dertinger2013advances}:
\begin{equation}
\label{eqn:cumulant_signal}
C_n(\bm r) = \sum_{i_1, \ldots, i_n} \left|h(\bm r - \bm s_{i_1})\right|^2\cdots \left|h(\bm r - \bm s_{i_n})\right|^2 w_{i_1 \ldots i_n},
\end{equation}
where $\bm s_i$ is the position of $i$th source; $w_{i_1 \ldots i_n} = \operatorname{cum}(\delta I_{i_1}, \ldots, \delta I_{I_n})$ is the cumulant of zero-mean processes, describing fluctuations of the sources \cite{mendel1991tutorial}; $\delta I_i(t) = I_i(t) - E(I_i)$, and $E(I)$ is the expectation value of a random process $I$. For example, for the 2nd and 4th order cumulants, one has
\begin{equation}
\label{eqn:cumulants_2}
w_{ij}=E(\delta I_i \delta I_j),
\end{equation}
\begin{multline}
\label{eqn:cumulants_4}
w_{ijkl}=E(\delta I_i \delta I_j \delta I_k \delta I_l) - E(\delta I_i \delta I_j) E(\delta I_k \delta I_l)\\
{}- E(\delta I_i \delta I_k) E(\delta I_j \delta I_l)- E(\delta I_i \delta I_l) E(\delta I_j \delta I_k).
\end{multline}

If the sources are independent, the expectation values factorize, $E(\delta I_i \delta I_j \cdots\delta I_k) = E(\delta I_i) E(\delta I_j) \cdots E(\delta I_k) = 0$ for $i \ne j \ne \ldots\ne k$, and the cumulants of the fluctuations of different sources vanish: $w_{i_1, i_2, \ldots, i_n} = 0$ unless $i_1 = i_2 = \ldots = i_n$.

In a real experiment, the expectation values are estimated as averages over finite-length data series, which are composed of intensities $I_i(t_j)$ integrated over finite number of frames $\{t_j\}$. I.e. each expectation value $E(X)$ is replaced by the average $\langle X \rangle = \sum_{j=1}^N X(t_j) / N$, where $N$ is the number of frames. 

\begin{figure}[tp]
    \centering
    \includegraphics[width=0.9\linewidth]{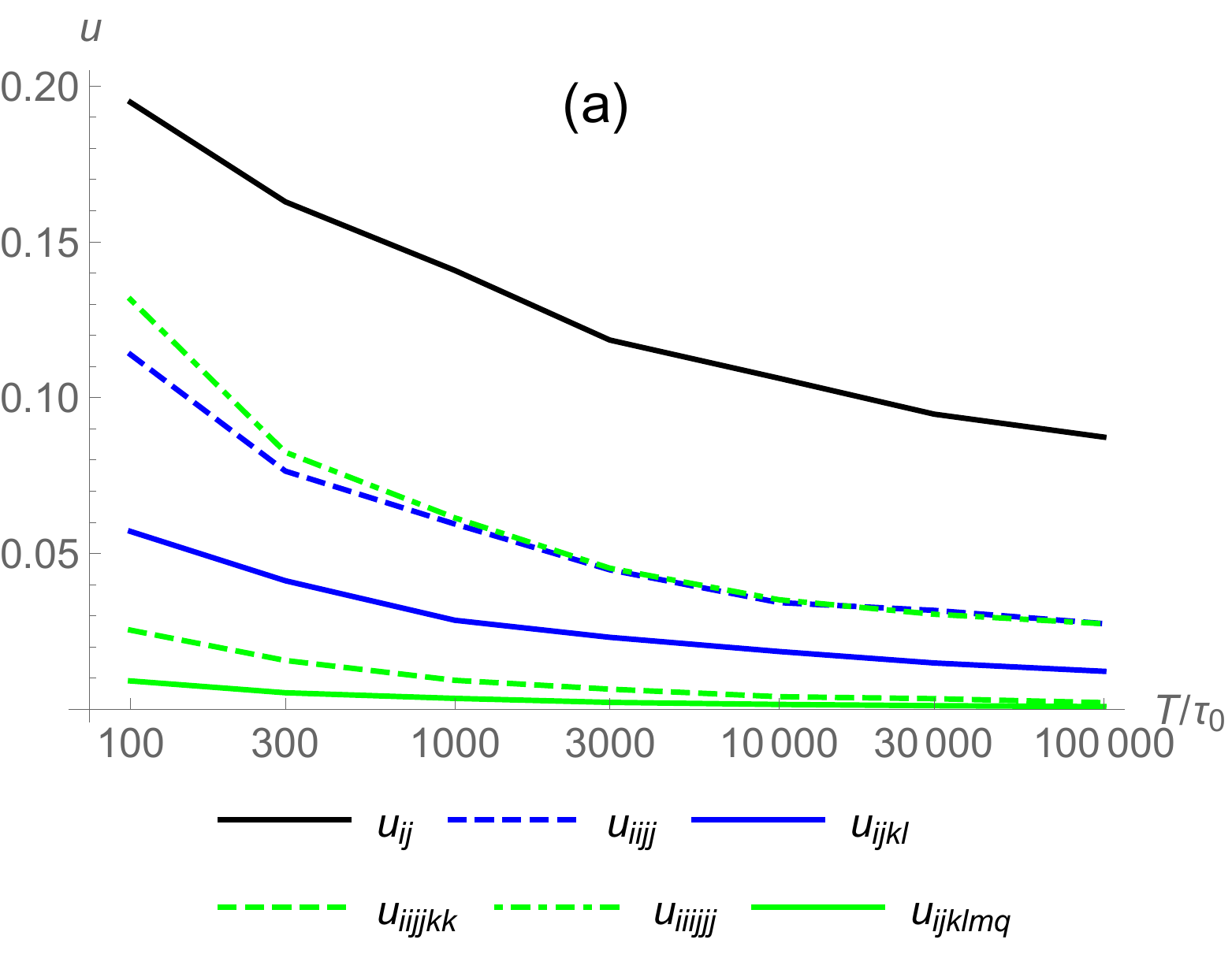}\\
    \includegraphics[width=0.9\linewidth]{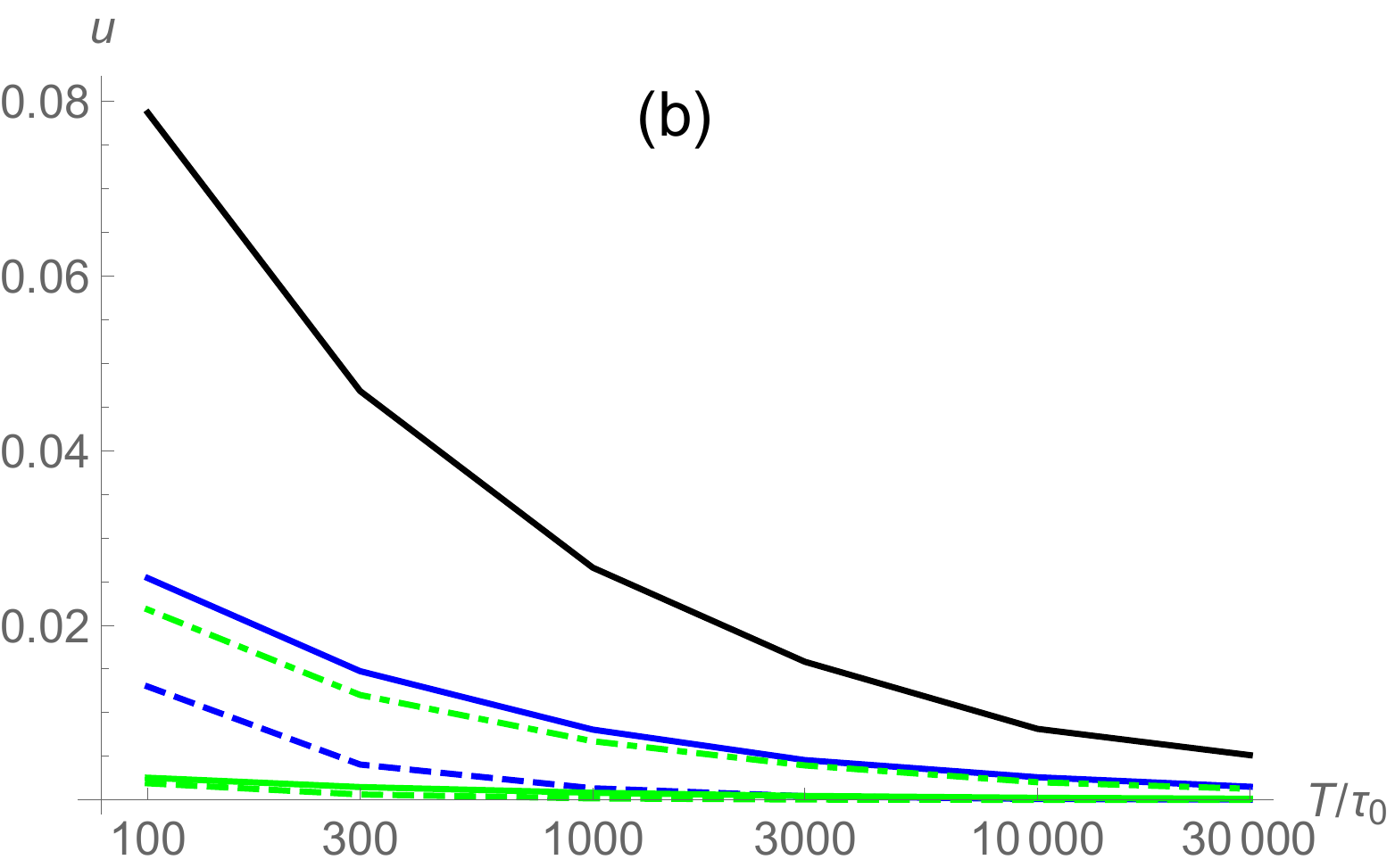}
    \caption{The dependence of the normalized 2nd (black), 4th (blue), and 6th (green) order cumulants (Eq.~(\ref{eqn:cumulant_ratio})) on the acquisition time $T$ divided by the characteristic switching time $\tau_0$ of the sources. The results are obtained by Monte-Carlo simulations. The values and variances of cumulant estimates in Eq.~(\ref{eqn:cumulant_ratio}) were calculated by modeling 1000 realization of the considered finite-length data series per each point in the plots. Dashed lines correspond to the cumulants with pairs of coinciding indices. Dot-dashed line describes the cumulant with two triples of coinciding indices. In the plot legend, $i\ne j \ne k \ne l \ne m \ne q$ is assumed. During modeling, the sources are assumed to be identical. The probability distribution of the switching times (both, for ``on'' and ``off'' states) was modeled by a power-law dependence $p(\tau) \sim (\tau / \tau_0)^{-\alpha}$ with $\alpha = 2$ (a) and 3 (b). The frame size is equal to $\tau_0$.}
    \label{fig:cumulants}
\end{figure}

Fig.~\ref{fig:cumulants} illustrates the difference between ideal cumulants $w_{i_1 \ldots i_n}$ and their estimates $\bar w_{i_1 \ldots i_n}$ over finite-length data series. The expectation values of the cumulant estimates still remain zero, $E(\bar w_{i_1 \ldots i_n}) = 0$, if at least two of the indices $i_1$, \ldots, $i_n$ are different. However, the actually obtained values fluctuate from realization to realization and have non-zero variance: $\operatorname{Var}(\bar w_{i_1 \ldots i_n}) > 0$. Therefore, the effect of cancelling out the contributions from several sources to $C_n(\bm r)$, used during the derivation of Eq.~(\ref{eqn:cumulant_ideal}), occurs for infinite acquisition time only, while for finite-time experiments all the cumulants $\bar w_{i_1 \ldots i_n}$ contribute to the final signal and reduce the resolution. To quantify the effect, we introduce the ratios
\begin{equation}
\label{eqn:cumulant_ratio}
u_{i_1\dots i_n} = \sqrt{\operatorname{Var} \bar w_{i_1 \ldots i_n}} / |\bar w_{i\dots i}|,
\end{equation}
describing the characteristic value of the joint contribution of the sources $i_1$, \ldots, $i_n$ to $C_n(\bm r)$, normalized by the contribution of a single source (here, we assume the sources to be identical). Fig.~\ref{fig:cumulants} shows the dependence of the constructed ratios for the 2nd, 4th, and 6th order cumulants on the acquisition time $T$, expressed in terms of the characteristic switching time $\tau_0$. One can see that the contribution of the cumulants, which are expected to have zero values, remains considerable even for $T / \tau_0 \sim 10^5$.

The effect of shot noise amplification during cumulants calculation is closely connected to the fact that the difference of two random variables with Poisson distributions with the mean values $\mu_1$ and $\mu_2$ is described by the Skellam distribution,  with the variance $\mu_1 + \mu_2$ being larger than the mean $\mu_1 - \mu_2$. For $\mu_1 = \mu_2$, the signals cancel in average, but yield twice as large variance ($\sqrt 2$ times larger shot noise). Fig.~\ref{fig:noise} of the main text illustrates the effect by showing the intensity, the 2nd, and the 4th order cumulants for two sources together with the error, caused by the shot noise. The 4th order cumulant, as expected, demonstrates better separation of the images of the sources. However, its fluctuations are also much stronger than the ones of lower-order cumulants.

\section{Simulations for different objects: different configurations of 2,3, and 4 sources, different brightness and source states} 

Here, we present the results of simulations, performed for different objects with 2, 3 and 4 sources in 1D and 2D configurations (see also Fig.~\ref{fig:configuration 2D}). 

\begin{figure*}[tp]
    \centering
    \includegraphics[width=0.9\textwidth]{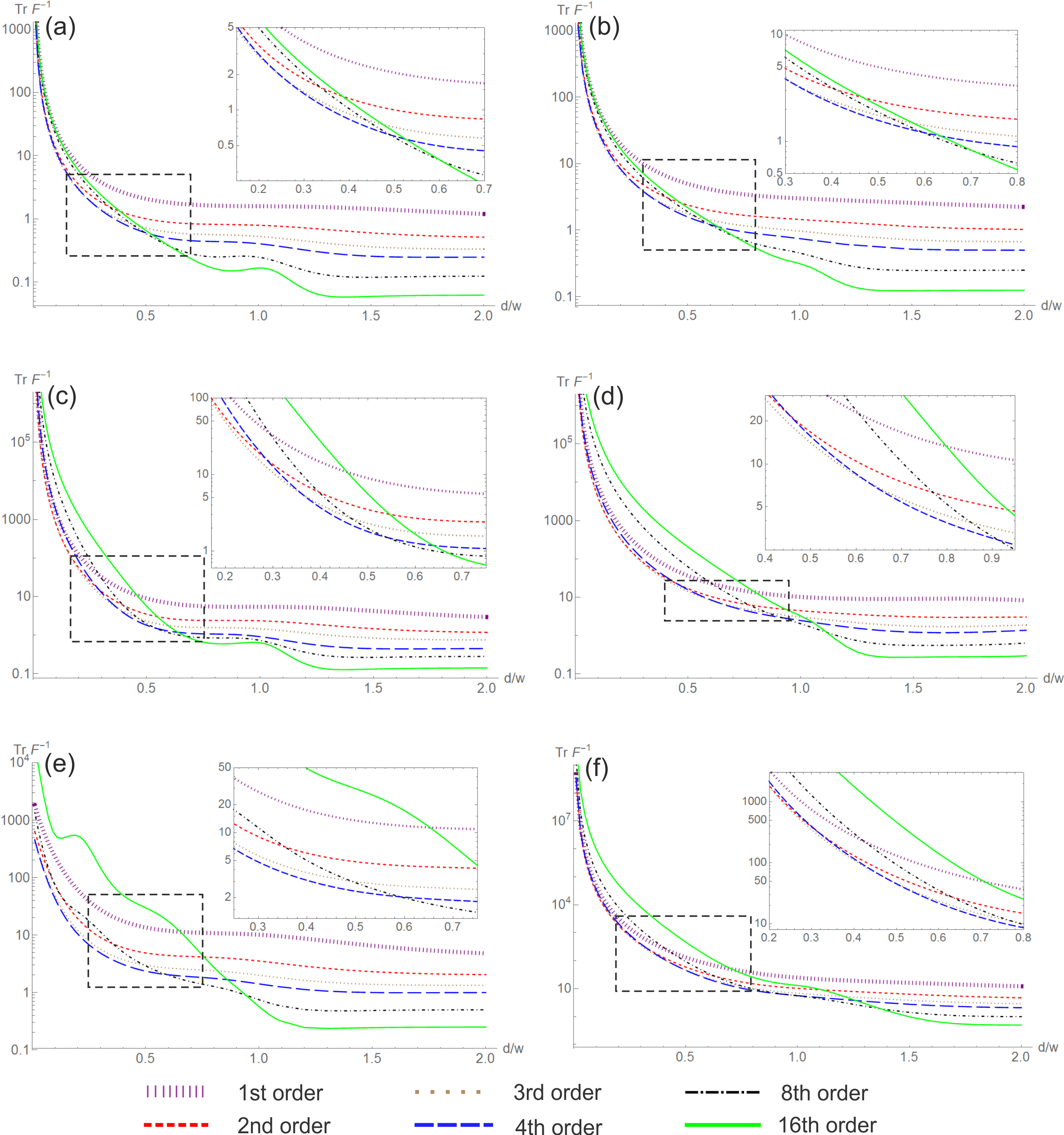}
    \caption{The dependence of the total reconstruction error, estimated as the trace of the inverse of Fisher information matrix, on the normalized distance between adjacent sources for 2 (a), (b) 3 (c), (d) and 4 (e), (f) sources in the 1-dimensional (left) and 2-dimensional (right) cases. The curves correspond to different orders of the analyzed correlation functions. The insets show enlarged parts of the plots with intersections of the curves. The parameters $\alpha = 0.3$ and $\xi = 0.4$ are the same for all the sources. The set of detection positions was taken the same as for Figs. \ref{fig:error vs scale} and \ref{fig:error vs order}.}
    \label{fig:curves full}
\end{figure*}

In Fig.~\ref{fig:curves full}, the trace of the inverse Fisher matrix is shown for 1D and 2D objects, consisting of 2, 3, and 4 sources, in the way similar to that for a 1D object with 3 sources  considered in the main text and shown in  Fig.~\ref{fig:error vs scale}. One can see that, despite different configurations and number of sources, there are common tendencies in all the considered cases. In the ``classical'' resolution regime (approximately, $d/w > 1$), measuring correlation functions of a higher order brings about increase of the information content and improvement of the resolution. However, even in the ``near super-resolution'' regime one observes the inverse situation  (see insets in Fig.~\ref{fig:curves full}).

\begin{figure*}[tp]
    \centering
    \includegraphics[width=\linewidth]{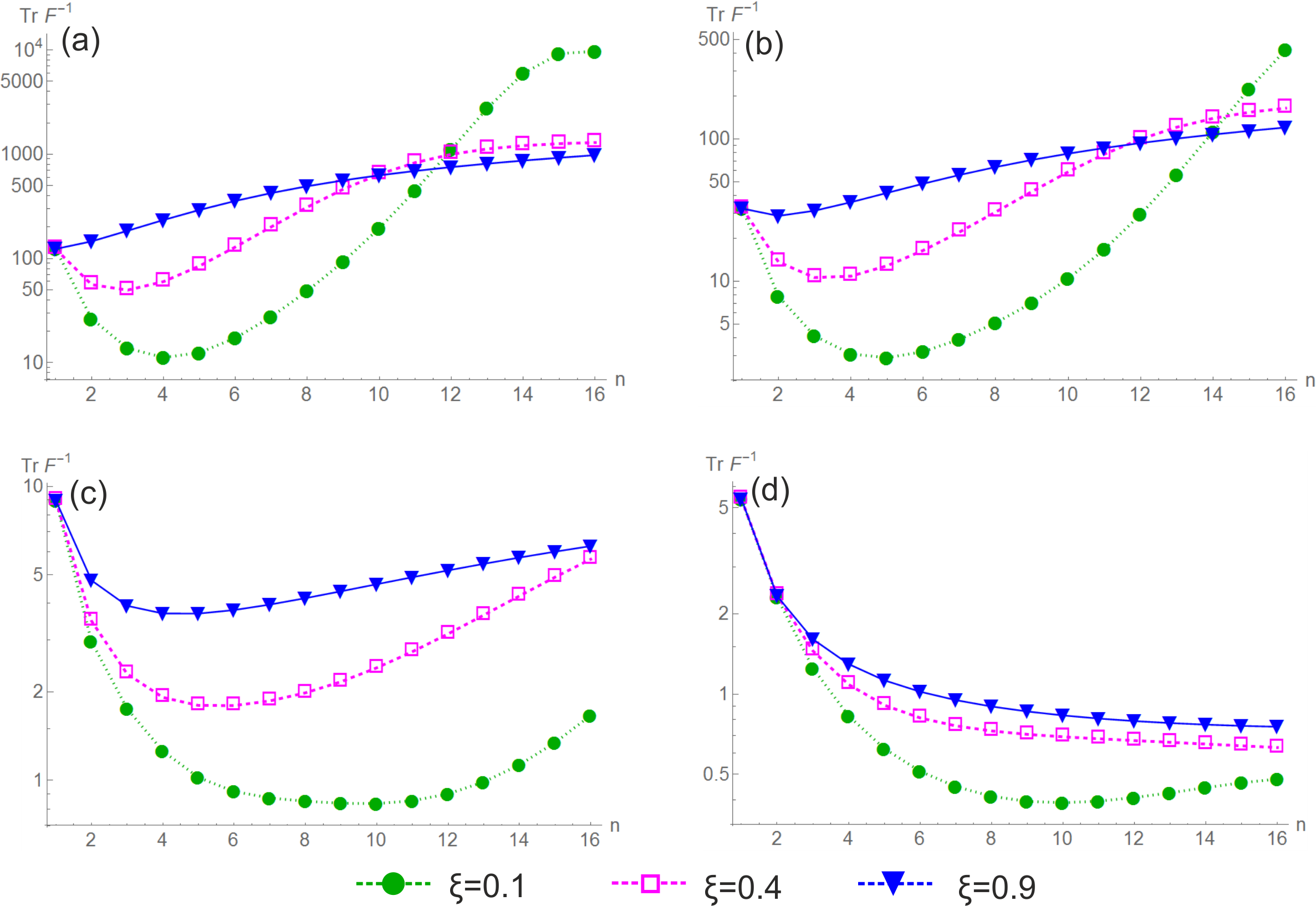}
    \caption{The dependence of the total reconstruction error on the order of the analyzed correlation function in the 1-dimensional case for 3 sources for several values of $\xi$ and different regimes of imaging: (a) $d/w=0.2$, (b) $d/w=0.3$, (c) $d/w=0.5$, (d) $d/w=1$. Other parameters are as for Fig.~\ref{fig:curves full}.}
    \label{fig:vert sections various xi}
\end{figure*}

\begin{figure*}[tp]
    \centering
    \includegraphics[width=\linewidth]{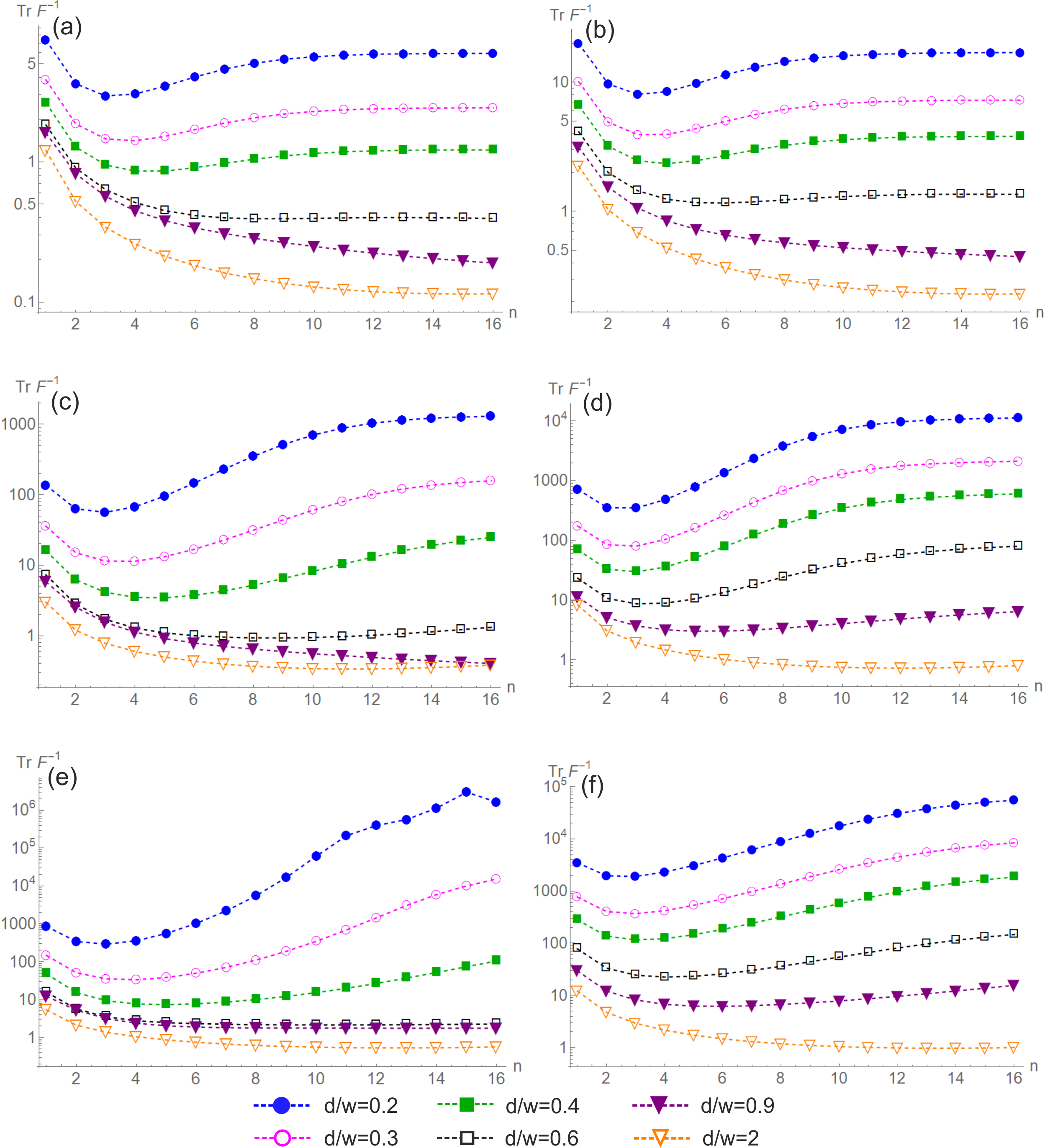}
    \caption{The dependence of the total reconstruction error on the order of the analyzed correlation function in the 1-dimensional (left) and 2-dimensional (right) cases for the objects, where the amplitudes $\alpha_i$ are different for the sources constituting the object: for 2 sources (a), (b) the amplitudes are $\alpha_i=\{0.285, 0.3\}$; for 3 sources (c), (d): $\alpha_i=\{0.285, 0.3,0.309\}$; for 4 sources (e), (f): $\alpha_i=\{0.285, 0.294,0.3,0.309\}$. The probability of the ``bright'' regime is $\xi = 0.4$. Other parameters are as for Fig.~\ref{fig:curves full}.}
    \label{fig:vert sections various alpha}
\end{figure*}

The values of the trace of the inverse Fisher matrix are shown for different probabilities of the ``bright'' state in Fig.~\ref{fig:vert sections various xi} and for different amplitudes of the sources within the same object  (Fig. \ref{fig:vert sections various alpha}). Generally, the already much discussed presence of the optimal correlation order is also observed in all the considered cases. An additional nontrivial feature is that for the sources with high brightness (i.e., when each source is in the ``bright'' state for much longer than in the ``dark'' state), the optimal correlation order might be the lowest one in the super-resolution regime (see Fig.~\ref{fig:vert sections various xi} (c-d)). This result can be understood from Eq. (\ref{eqn:Gn_2sources}): the relative contribution of the terms with $m = 1$, \ldots, $n-1$ in the expression for $C^{(n)}$ increases with the growth of $\xi$, thus decreasing the contrast of the image. For such cases, the SOFI is practically irrelevant: the best results can be obtained by traditional intensity measurements.   

\end{document}